# Composition dependence of tracer diffusion coefficients in Fe–Ga alloys: a case study by a tracer-interdiffusion couple method

G.M. Muralikrishna[1,2], B. Tas[1], N. Esakkiraja[3], V. Esin[4], K.C. Hari Kumar[2], I.S. Golovin[5], I.V. Belova[6], G.E. Murch[6], A. Paul[3], S.V. Divinski[1,7]

[1] *Institute of Materials Physics, University of Münster, Germany*
[2] *Department of Metallurgical and Materials Engineering, IIT Madras, Chennai 600036, India*
[3] *Department of Materials Engineering, Indian Institute of Science, Bangalore 560012, India*
[4] *MINES ParisTech, PSL University, Centre des Matériaux (CMAT), CNRS UMR 7633, Évry, France*
[5] *National University of Science and Technology "MISIS", Leninsky ave. 4, 119049, Moscow, Russia*
[6] *University Centre for Mass and Thermal Transport in Engineering Materials, Priority Research Centre for Geotechnical and Materials Modelling, School of Engineering, The University of Newcastle, Callaghan, NSW 2308, Australia, Australia*
[7] *Samara National Research University, Moskovskoye Shosse 34, Samara 443086, Russia*

The problem of estimation of the tracer diffusion coefficients is solved by utilizing a novel tracer-interdiffusion couple technique even in the absence suitable radioisotope of one of the components and absence of reliable thermodynamic parameters. This is demonstrated by generating reliable and reproducible mobility data in the alloys of the Fe–Ga system with a strong composition dependence of the diffusion coefficients. Tracer- ($^{59}$Fe) and inter-diffusion are simultaneously measured in three couples Fe/Fe-16Ga, Fe/Fe-24Ga and Fe-16Ga/Fe-24Ga at 1143 K. The results obtained for the couples with different end-members are in an excellent agreement with each other for the overlapping composition intervals. The influence of the molar volume on the measured tracer- ($^{59}$Fe) and inter-diffusion coefficients is evaluated. Using thermodynamic calculations, the Ga tracer diffusion coefficient and the vacancy wind factor are determined via the Darken-Manning relation for the composition range of 0-24 at.% Ga. The results confirm the reliability of the tracer-interdiffusion couple technique for producing highly accurate diffusion data, in the present case for optimizing the mobility description of the bcc phase of the Fe-Ga system. The Ga tracer diffusion coefficients are further estimated via experimental determination of the ratio of the Fe and Ga tracer diffusivities at the Kirkendall marker planes and utilizing the Fe tracer diffusion coefficients measured directly by the radiotracer method.

## 1. Introduction

Tracer and chemical (or inter-) diffusion represent two complementary, but basically different approaches for measuring long-range atomic transport in solids. Tracer diffusion is measured using tiny amounts of 'marked' (typically radioactive or highly enriched stable) isotopes in an alloy of a given composition, thus under a purely entropic driving force. The tracer diffusion coefficient of an element A, $D_A^*$, can conveniently be determined by applying the known solution of the diffusion problem to the measured concentration profiles [1–3]. Obviously, in order to determine the concentration dependence of $D_A^*$, a large number of independent measurements in different (homogeneous) alloys is required and the number of required compositions would scale according to a power law with the number of components $n$. Alternatively, chemical (or inter-) diffusion is measured using two (or more) alloys of different compositions brought into contact. In this case, the diffusion transport is induced by the gradients of the chemical potentials. In the binary case, the interdiffusion coefficient, $\widetilde{D}$, can conveniently be determined using Boltzmann-Matano method [4] using, for example, Sauer-Freise formalism [5] for the composition interval between the two end-members [6]. In the ternary case, the Matano-Kirkaldy method [7] can be used to determine the matrix of interdiffusion coefficients, $\widetilde{D}_{ij}^n$, ($i,j = 1,2$ and $n = 3$ is the reference element), but only at the composition of intersection of the two independent diffusion paths. For $n$-component alloys ($n \geq 4$) there is generally no chance to apply this method established on the well-known Onsager formalism [8,9] to estimate the whole $(n-1) \times (n-1)$ matrix of the independent interdiffusion coefficients $\widetilde{D}_{ij}^n$ ($i,j = 1,..,n$-1), since the given number of independent diffusion paths (by definition, one dimensional) cannot be forced to intersect in a multicomponent space (the $n^{\text{th}}$ element is chosen as the reference one here).

Although diffusion in multicomponent alloys is of high technological relevance, its analysis, therefore, becomes cumbersome if the number of components exceeds three [10]. In order to resolve this standoff situation, several approaches have been elaborated recently. The body-diagonal diffusion couple method suggested by Morral [11] allows determining the full matrix of the interdiffusion coefficients in a relatively narrow composition interval of constant diffusivities in multicomponent alloys. It was successfully applied to the high-entropy CoCrFeNi and CoCrFeMnNi systems by Verma *et al.* [12] after compromising the strict condition of intersecting the diffusion paths since they noted that diffusion couples do not intersect even in such a small composition range of more or less constant diffusivities. Alternatively, Paul with co-workers developed pseudo-binary and pseudo-ternary approaches [13,14] which provides the interdiffusion coefficients and could be used to determine the tracer diffusion coefficients using a proper thermodynamic description, if available [15]. However, the latter methods are useful only when two (in a pseudo-binary diffusion couple) or three (in a pseudo-ternary diffusion couple) components develop the diffusion profiles keeping all other components constant in the interdiffusion zone in an $n$-component system [16].

Alternatively, a tracer-interdiffusion couple technique proposed recently [17–19] provides potentially a strong tool for producing basic kinetic data, which could be used as input to create CALPHAD-type mobility databases [19]. This type of experiments corresponds to measurements of tracer diffusion under a chemical gradient in alloys to evaluate the concentration-dependent tracer diffusion coefficients as it was suggested by Manning [20]. In [21] a novel tracer-interdiffusion couple technique was applied to the original diffusion dataset of Manning [20,22] for the Ag–Cd system. The tracer profiles from [20,22] were re-evaluated, and consistent tracer diffusion data were obtained [21]. That system is a very convenient one, offering suitable radioisotopes for both elements, Ag and Cd. However, the partial molar volumes in the Ag-Cd system are almost equal and practically do not change along the diffusion path.

The present paper aims to investigate tracer and interdiffusion in Fe–Ga alloys applying the tracer-interdiffusion couple technique with a sub-goal to test the approach for a system with a strong variation of the molar volume and an expected significant change of the diffusion coefficients. Fe–Ga alloys have attracted attention due to their giant magnetostriction in low magnetic fields [23–25] and a number of first- and second-order phase transitions below ~1000 K [26,27]. The Fe–Ga alloys reveal distinct ordering and the type of order depends on temperature and concentration of Ga atoms [25,28]. To the best of our knowledge, no tracer diffusion data have been reported for the Fe-Ga alloys so far, and the diffusion properties of the Fe–Ga alloys are essentially unknown in spite of their key role for diffusion-controlled phase transitions. While Fe diffusion can straightforwardly be measured using the radiotracer technique and applying the relatively long-lived (half-life of 45 d) $^{59}$Fe isotope [29], Ga offers only two suitable and relatively short-living radioisotopes, $^{67}$Ga (half-life of 3 d) and $^{72}$Ga (half-life of 14 h) that hinders reliable diffusion measurements especially at moderate and lower temperatures. As discussed in this article, we solve this issue for the estimation of tracer diffusion coefficients of both the components at a relatively high temperature of 1143 K in a single-phase region of the binary Fe–Ga system [30] by following the tracer-interdiffusion couple technique.

## 2. Materials and Methods
### 2.1. Alloy preparation

High purity Fe (99.96 wt.%) and Ga (99.99 wt.%) elements were used to synthesize Fe-16at.%Ga and Fe-24at.%Ga (hereafter, denoted also as 16Ga and 24Ga, respectively) alloys using induction melting under a high-purity argon atmosphere. The ingots were homogenized by annealing at 1373 K for two days. The homogenized ingots were cut into 1 mm thick discs of 8 mm in diameter by spark erosion. The samples were polished to a mirror-like finish following the standard metallographic sample preparation procedures. Polished samples were subjected to a pre-annealing treatment at 1143 K for 24 h to ensure a near-equilibrium structure for subsequent diffusion measurements. Furthermore, this pre-annealing treatment helped to remove any polishing-induced stresses and to attain equilibrium before diffusion experiments at the same temperature. A similar preparation was done for the samples of pure Fe. The chemical composition of pure Fe is given in Table 1.

**Table 1**. Chemical composition of the Fe sample (in wt. ppm). Fe amount is balanced.

| C | N | O | Al | Si | P | S | Cr | Co | Ni | Cu | Zn |
|---|---|---|----|----|---|---|----|----|----|----|----|
| 25 | 96 | 4 | <2 | <50 | <20 | <50 | 4 | <5 | <5 | 900 | 170 |

The chemical composition of the homogenized samples was examined using energy dispersive spectroscopy (EDS) attached to a scanning electron microscope (SEM, FEI Nova NanoSEM 230).

The alloys and pieces of pure Fe were used to assemble couples for the combined tracer-interdiffusion experiments. Due to obvious limitations in experiments involving open radioactive substances, two sets of identical couples were prepared, one for purely interdiffusion experiments using electron probe micro-analysis (EPMA) and one for the tracer measurements under a concentration gradient.

### 2.2. Interdiffusion experiments

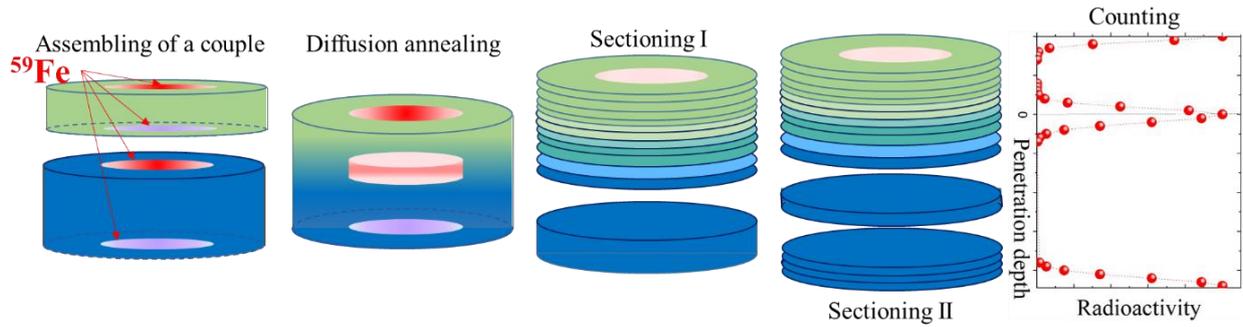

**Figure. 1.** Main steps of the tracer-interdiffusion couple experiment: (from left to right) tracer application; assembling of the couple and diffusion annealing; parallel sectioning starting from the thinner specimen side in the bonded couple which is continuing after reaching of the Matano plane till background is approached; sectioning from the opposite (thicker) specimen side till background is reached; activity counting and plotting the full concentration profile.

The pre-annealed samples of Fe-16Ga and Fe-24Ga alloys and pure Fe were sandwiched to produce the diffusion couples:
- Diffusion couple 1 (DC 1): Fe-16Ga / Fe-24Ga,
- Diffusion Couple 2 (DC 2): Fe / Fe-16Ga and
- Diffusion Couple 3 (DC 3) Fe / Fe-24Ga.

The sandwiched couples were assembled in custom-made mechanical fixtures made of heat-resistant THERMAX steel. Tantalum spacers were used to avoid any contact between the fixtures and the diffusion couples. The screws of the fixture were just hand-tightened after placing the diffusion couples to avoid imposing any external stresses. The assembled diffusion couples were placed in quartz tubes, evacuated to a $6\times10^{-3}$ Pa residual pressure, then filled with high purity Ar and finally sealed. The samples were diffusion annealed at 1143 K for 5 h. The furnace temperature was well-calibrated using a Ni-NiCr thermocouple to an accuracy of ±1 K.

Annealed interdiffusion couples were cross-sectioned with a slow-speed diamond saw and embedded in epoxy. The couple halves were then ground and polished for the diffusion profile measurements in an electron probe micro-analyzer (EPMA, CAMECA SX100) using 15 kV, 40 nA at a step size of 1 µm. Pure components were used as standards. Three profiles were measured in each diffusion couple to make sure of the consistency profiles considered for estimation of the data. All the profiles in a particular diffusion couple are found to be very similar. Moreover, as explained in the results and discussion sections below, a good agreement of the data estimated from different incremental diffusion couples indicates a good quality of the diffusion couples produced in this study.

### 2.3. Combined tracer-interdiffusion couple experiments

The samples prepared in identical conditions to the interdiffusion experiments were used for combined tracer-interdiffusion measurements. The $^{59}$Fe radioactive isotope with the activity of about 5 kBq was deposited on both surfaces of the Fe-16Ga, Fe-24Ga and Fe samples using the drop-and-dry technique. The deposited samples were arranged in a special fixture and subjected to diffusion annealing following the procedure identical to the interdiffusion experiments. As an additional benefit, this arrangement allows to measure the tracer diffusion coefficients in the unaffected end-members (these conditions correspond to standard radiotracer experiments) and the tracer diffusion coefficients in the interdiffusion zone along the diffusion path on both sides of the Matano plane simultaneously.

After diffusion annealing, the radioactive couples were reduced in diameter by about 1 mm to avoid the influence of artefacts from the surface and/or lateral diffusion. A precision parallel sectioning technique via mechanical grinding was used to obtain the penetration profiles of the $^{59}$Fe isotope. The experimental procedure is sketched in Fig. 1. Each diffusion couple was sectioned from one end to another, starting from a one end-member (prepared as a thinner disk) to the interdiffusion zone and continuing further till the background for the $^{59}$Fe isotope was safely reached (Sectioning I in Fig. 1). Then the sample was flipped, and the sectioning was continued from another end-member (prepared as a thicker disk), again till the background was approached (Sectioning II in Fig. 1). This procedure of sectioning facilitated the accurate measurements of the penetration profiles for unaffected end-members where tracer diffusion proceeded under purely an entropic driving force, see upper and bottom parts of the penetration profile sketched in Fig. 1, right panel. The sectioning procedure also allowed to measure the penetration profiles on both sides of the Matano plane which corresponded to tracer diffusion under a chemical driving force, see the central part of the penetration profile sketched in Fig. 1, right panel. An intrinsic Ge γ-detector equipped with a 16K multi-channel analyzer was used to count the radioactive decay of the $^{59}$Fe γ-isotope for each section. The weight of the sample after each section being removed was measured on a microbalance to a relative accuracy of 0.1 μg and the section thickness was determined accounting for the change of the material density with the penetration depth. The local density was estimated after adjusting the tracer penetration profiles with the interdiffusion profiles. To this end, the origin of the abscissa for the tracer profile was set to the section which reveals maximum activity in the central part of the profile, black-dashed line in Fig. 1, right panel, and this position was assumed to correspond to the Matano plane determined from the chemical profile. Before the experiment, the total thickness of the couple was carefully measured to provide a continuous depth coordinate.

The density variation due to the chemical diffusion was taken into consideration by estimating the variation in the molar volumes. This was done by making use of the measured lattice parameters by X-ray diffraction.

### 3. Results and Discussion
### 3.1. Interdiffusion experiments

The relations for estimating the interdiffusion coefficients, $\widetilde{D}$, considering the actual molar volume variation proposed by Wagner [31] and den Broeder [32] are generally used for estimation of the interdiffusion coefficients. The Wagner relation [33] is expressed as:

$$\widetilde{D}(Y_B^*) = \frac{V_m^*}{2t}\left(\frac{dx}{dY_B}\right)_{x^*}\left[(1-Y_B^*)\int_{x^{-\infty}}^{x^*}\frac{Y_B}{V_m}dx + Y_B^*\int_{x^*}^{x^{+\infty}}\frac{1-Y_B}{V_m}dx\right] \quad (1)$$

where $Y_B = \frac{N_B - N_B^-}{N_B^+ - N_B^-}$ is the composition normalized Sauer-Freise variable [5], $N_B$ is the mole fraction of the component B, $N_B^-$ ($N_B^+$) is the composition corresponding to the left (right) unaffected end-member, $V_m$ is the molar volume, $t$ is the annealing time, and $x$ is the coordinate perpendicularly to the original interface. The asterisk indicates the position or composition of interest for the estimation of the interdiffusion coefficient.

The Den Broeder equation [32] is expressed as

$$\widetilde{D}(C_B^*) = \frac{1}{2t}\left(\frac{dx}{dY_C}\right)_{x^*}\left[(1-Y_C^*)\int_{x^{-\infty}}^{x^*}Y_C\,dx + Y_C^*\int_{x^*}^{x^{+\infty}}(1-Y_C)\,dx\right] \quad (2)$$

where $Y_C = \frac{C_B - C_B^-}{C_B^+ - C_B^-}$ is a normalized variable and $C_B = \frac{N_B}{V_m}$ is the concentration.

For a constant molar volume (leading to $Y_C = Y_B$) both the relations are transformed to the same equation, which can be expressed as

$$\widetilde{D}(Y_B^*) = \frac{1}{2t}\left(\frac{dx}{dY_B}\right)_{x^*}\left[(1-Y_B^*)\int_{x^{-\infty}}^{x^*}Y_B\,dx + Y_B^*\int_{x^*}^{x^{+\infty}}(1-Y_B)\,dx\right] \quad (3)$$

It should be noted here that the interdiffusion coefficients considering a constant molar volume or a linear variation between the molar volumes of the end-member compositions will give the same value because of the nature of the equations. This linear variation between the pure components as the end-members follows Vegard's law. However, it is different in an incremental diffusion couple in which molar volumes of the end-members are considered for the linear variation, as shown in Fig. 2.

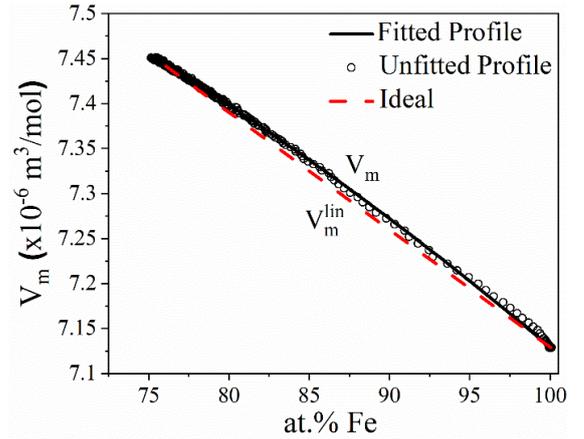

**Figure 2.** Molar volume variation in the system Fe–Ga as a function of the atomic percentage of Fe. The dashed line represents an ideal variation according to Vegard's law. The solid line corresponds to the parabolic fit, Eq. (4).

Wagner and Den Broeder derived the relations differently to arrive at equivalent relations. Baheti and Paul have shown the equivalence of these methods recently by deriving the Den Broeder relation following Wagner's approach [34]. However, for the actual non-ideal molar volume variation, there can be a small difference in the estimated diffusion coefficients depending on the extent of non-ideality because of a different level of errors accumulated during different steps followed in these two methods. This issue is analyzed here for the Fe–Ga system. Figure 2 shows the variation of the calculated molar volume, obtained from the experimentally determined lattice parameters $a$, with the alloy composition. In the composition range, the lattice parameter of the bcc unit cell, $a$ in Å, is best fitted by:

$$a = 2.8714 + 0.2023 \cdot (1 - N_{Fe}) - 0.1211 \cdot (1 - N_{Fe})^2 \qquad (4)$$

The positive deviation of the molar volume is evident in reference to the linear variation of the molar volume $V_m^{lin}$, as shown by the red dashed line in Fig. 2.

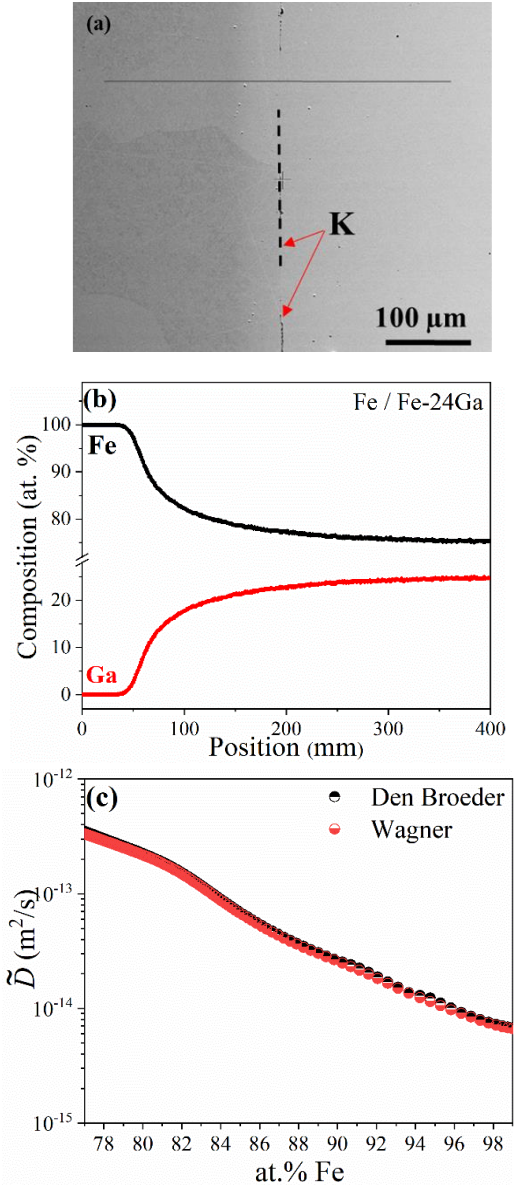

**Figure 3.** (a) SEM image showing the microstructure of the Fe/Fe-24Ga diffusion couple annealed at 1143 K for 5 hours. 'K' is the positoin of the Kirkendall marker plane. (b) Example of chemical profiles of constituent elements in the diffusion couple measured using EPMA. (c) Interdiffusion coefficients estimated in Fe/Fe-24Ga diffusion couple following the Wagner and Den Broeder methods.

Figures 3 a,b show the SEM image and the measured composition profile of the Fe/Fe-24Ga diffusion couple, respectively. The estimated interdiffusion coefficients following the Wagner and Den Broeder methods are shown in Fig. 3c. The difference in the estimated data following these two methods is found to be negligible.

Now we estimate the interdiffusion coefficients considering the actual molar volume variation following Eqs. (1) or (2) and considering a constant molar volume following Eq. (3). The positive deviation of the molar volume leads to an expansion of the diffusion couple. Considering a linear expansion, we can calculate this from [35]:

$$\pm \Delta x = \pm \int_{x-\infty}^{x+\infty} \frac{\Delta V_m}{V_m} dx \qquad (5)$$

Here $\Delta V_m = V_m - V_m^{lin}$ is the deviation of the actual molar volume from the ideal behaviour. The plus and minus signs correspond to the positive and negative deviation leading to expansion or shrinkage of the diffusion couple. It should be noted here that $\Delta x$ is also equal to the difference in location of the Matano planes when calculated considering the concentration profiles of Fe and Ga because of non-ideality of the molar volume variation [35]. Our analysis for the Fe/Fe-24Ga diffusion couple with the interdiffusion zone length ~175 μm indicates that we have an expansion of only 0.2 μm due to such a small positive deviation of molar volume in this system.

We further discuss an important aspect of measuring the composition profile of a diffusion couple far enough towards unaffected end-members, which can be considered as a textbook example. In Fig. 5, two profiles measured on the same diffusion couple Fe/Fe-24Ga are shown. The composition profile in Fig. 5a (profile 1) is not measured correctly up to the unaffected part of the Fe-24Ga alloy. The correct measurement covering the whole interdiffusion zone is shown in Fig. 5a (profile 2). The difference in estimated interdiffusion coefficients from these two profiles can be explained by rewriting Equation 1 as:

$$\widetilde{D}(Y_B^*) = \frac{V_m^*}{2t} \frac{dx}{dN_B} (N_B^+ - N_B^-) \left[ (1 - Y_B^*) \int_{x-\infty}^{x^*} \frac{Y_B}{V_m} dx + Y_B^* \int_{x^*}^{x+\infty} \frac{1-Y_B}{V_m} dx \right] \qquad (6)$$

The first term, $\frac{V_m^*}{2t} \frac{dx}{dN_B}$, is the same for the two profiles at a given composition. However, the residual terms in Equation 6 will yield a lower value for the incomplete profile 1 because of a smaller value of $(N_B^+ - N_B^-)$ and a similar effect for the term in the brackets. Thus, a systematically smaller interdiffusion coefficient is determined for all concentrations because of the nature of the equation. This underlines the importance of measuring the diffusion profiles correctly for an accurate estimation of the diffusion coefficients.

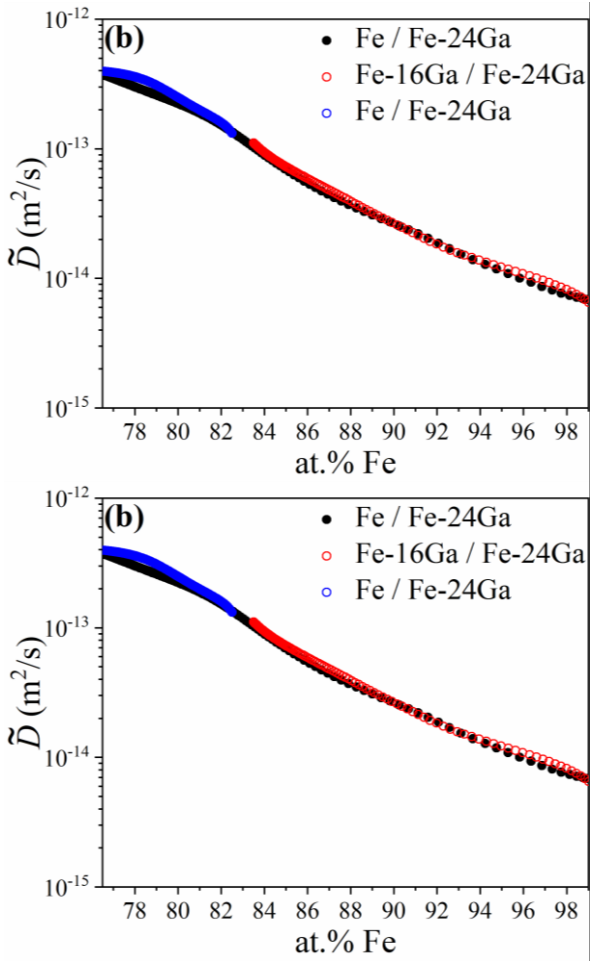
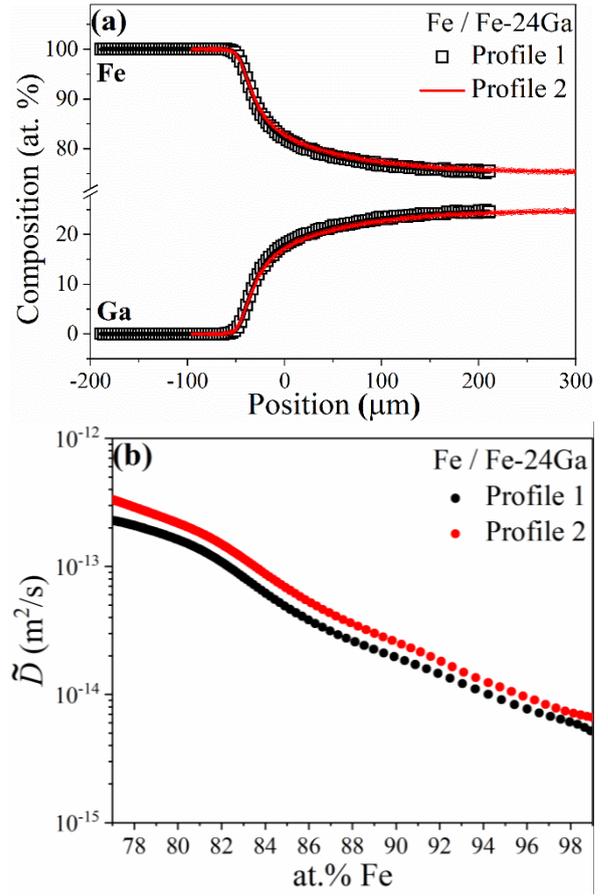

**Figure 4.** (a) Compiled plot of the interdiffusion diffusion coefficients at 1143 K with the variation of Fe concentration obtained from Fe-16Ga/ Fe-24Ga, Fe/ Fe-16Ga, and Fe/ Fe-24Ga diffusion couples estimated by Equation (1) considering the actual molar volume variation (b) the interdiffusion coefficients calculated from these couples estimated by Equation 3 considering a constant molar volume.

Therefore, we expect only a minor difference in the interdiffusion coefficients considering either the actual molar volume variation or a constant molar volume. We have estimated these using Eqs. (1) and (3) in all the diffusion couples, as shown in Fig. 4. The minor difference in results is evident because of the very small deviation of the molar volume in this system. A higher difference can be found in other systems depending on the extent of departure of the molar volume from the ideality [34]. It also should be noted here the similarities of interdiffusion coefficients when estimated from different incremental diffusion couples. Considering a higher difference for the data reported previously in Ni-Pd [36] and Ni-Al systems [6], we can state that the quality of the diffusion couples produced in this system is excellent.

**Figure 5.** Importance of measuring the extended composition profiles of the end-members: (a) Profile 1 was measured symmetrically with respect to the Matano plane but did not cover the whole interdiffusion zone, and Profile 2 was measured according to the established changes of the composition covering the whole interdiffusion zone. (b) The estimated interdiffusion coefficients at 1143 K, for 5 h.

The ratio of the intrinsic diffusion coefficients can be estimated at the Kirkendall marker plane by [37]:

$$\frac{D_B}{D_A} = \frac{\bar{V}_B}{\bar{V}_A} \frac{\left[N_B^+ \int_{x-\infty}^{x^K} Y_{N_B} dx - N_B^- \int_{x^K}^{+\infty}(1-Y_{N_B})dx\right]}{\left[-N_A^+ \int_{x-\infty}^{x^K} Y_{N_B} dx + N_A^- \int_{x^K}^{+\infty}(1-Y_{N_B})dx\right]} \quad (7)$$

The interdiffusion, intrinsic and tracer diffusion coefficients are related by the Darken-Manning equation [38,39]:

$$\tilde{D} = C_B V_B D_A + C_A V_A D_B \quad (8a)$$

$$D_B = \frac{V_m}{\bar{V}_A} D_B^* \Phi(1+W_B) \quad (8b)$$

$$D_A = \frac{V_m}{\bar{V}_B} D_A^* \Phi(1-W_A) \quad (8c)$$

where the terms $W_i = \frac{2N_i(D_A^* - D_B^*)}{M_0(N_A D_A^* + N_B D_B^*)}$ represents the vacancy–wind effects, $M_0$ is a constant and depends on the crystal structure (5.33 in the BCC solid solution phase of interest). $\Phi = \dfrac{d\ln a_A}{d\ln N_A} = \dfrac{d\ln a_B}{d\ln N_B}$ is the thermodynamic factor, which is the same for both components A and B in a binary system due to the Gibbs-Duhem relation and $a_i$ is the activity of component $i$.

Table 2. Estimated ratios of intrinsic, $D_{Ga}/D_{Fe}$, and tracer, $D^*_{Ga}/D^*_{Fe}$, diffusion coefficients at the Kirkendall plane positions in the FeGa couples at 1143 K along with other parameters in Eqs. (8a)-(8c).

| Couple | $V_m$ | Composition (K plane) $N_{Fe}$ (at. %) | $D$ ($10^{-13}$ m²/s) | | | $\dfrac{D_{Ga}}{D_{Fe}}$ | $1+W_{Ga}$ | $1-W_{Fe}$ | $\left(\dfrac{D^*_{Ga}}{D^*_{Fe}}\right)$ | $D^*_{Fe}$ | $D^*_{Ga}$ | $\Phi$ |
|---|---|---|---|---|---|---|---|---|---|---|---|---|
| | | | $\widetilde{D}_k$ | $D_{Ga}$ | $D_{Fe}$ | | | | | ($10^{-13}$ m²/s) | | |
| Fe/Fe-24Ga | Constant | 82.8 | 1.16 | 1.19 | 1.02 | 1.17 | 1.01 | 0.96 | 1.12 | 0.35 | 0.39 | 3.0 |
| | Actual | | 1.15 | 1.25 | 0.83 | 1.51 | 1.02 | 0.91 | 1.33 | 0.37 | 0.49 | 2.4 |
| Fe-16Ga / Fe-24Ga | Constant | 78.7 | 3.22 | 3.65 | 1.65 | 2.21 | 1.05 | 0.82 | 1.72 | 0.84 | 1.45 | 2.4 |
| | Actual | | 3.21 | 3.71 | 1.49 | 2.49 | 1.06 | 0.97 | 1.85 | 0.81 | 1.50 | 2.3 |

From Equations 7 and 8, we have

$$\dfrac{D^*_B(1+W_B)}{D^*_A(1-W_A)} = \dfrac{D_B \bar{V}_A}{D_A \bar{V}_B} = \dfrac{\left[N_B^+ \int_{x-\infty}^{x^K} Y_{N_B} dx - N_B^- \int_{x^K}^{x+\infty}(1-Y_{N_B})dx\right]}{\left[-N_A^+ \int_{x-\infty}^{x^K} Y_{N_B} dx + N_A^- \int_{x^K}^{x+\infty}(1-Y_{N_B})dx\right]} \quad (9)$$

The marker plane positions could be detected in the Fe-24Ga and Fe16Ga-Fe24Ga couples. These are located at $N_{Fe} = 0.83$ and 0.79, respectively. We have estimated the ratio of the intrinsic diffusivities at these planes from Eq. (7). Together with the obtained interdiffusion coefficient, the individual values of the intrinsic diffusion coefficients can be determined from Eq. (8a). The estimated data are listed in Table 2. To examine the role of molar volume, we have estimated these data considering a constant molar volume (such that $\bar{V}_{Fe} = \bar{V}_{Ga} = V_m$) at the composition of interest and also by considering the actual variation of the molar volume. The partial molar volumes at these compositions are given in Table 2. A slightly higher difference is found in the estimated ratio of diffusivities (compared to the estimated interdiffusion coefficients) because of the ratio of partial molar volumes in Eq. 7. We can estimate the ratio of tracer diffusion coefficients considering Eq. (9) at the Kirkendall marker plane directly from the composition profiles. The estimated Fe tracer diffusion coefficients measured by the radiotracer method (as explained below) can then be used to determine the Ga tracer diffusion coefficients. Therefore, we do not need the details of thermodynamic data if not available. Furthermore, Eqs. (8b) and (8c) allow to determine the thermodynamic factors, which are listed in Table 2.

### 3.2. Radiotracer experiments

Figure 6 shows the penetration profiles of $^{59}$Fe measured for the combined tracer-interdiffusion couple experiments at 1143 K for the same annealing time of 5 h.

The annealing conditions at the unaffected end-members (terminal compositions) corresponded to the thin film geometry and Fig. 6 substantiates that the tracer penetration profiles, $C^*(x,t)$, for the end-members do follow the Gaussian solution of the diffusion problem:

$$C^*(x,t) = \dfrac{M}{\sqrt{\pi D_v t}} \exp\left(-\dfrac{(x-x_0)^2}{4 D_v t}\right) \quad (10)$$

where $M$ is the initial amount of tracer, $x$ is the penetration depth, $D_v$ is the tracer diffusion coefficient, and $x_0$ corresponds to the position of the outer surface (tracer origin). The $x$ coordinate for the tracer measurements in the interdiffusion zone was shifted in a way that its origin coincides with the position of the Matano plane, $x_M$, determined from the Bolzman-Matano analysis by:

$$x_M = \dfrac{\int_{C_i^-}^{C_i^+} x dC_i}{\int_{C_i^-}^{C_i^+} dC_i} = \dfrac{1}{C_i^+ - C_i^-} \int_{C_i^-}^{C_i^+} x dC_i \quad (11)$$

The tracer profiles, $C^*(x,t)$, for the end-members indicate distinct contributions of grain boundary diffusion at large depths, Fig. 6. These short-circuit contributions are neglected in the present analysis, and we are focused on volume tracer diffusion exclusively.

The impact of chemical gradients on tracer diffusion is explicitly featured by the strongly asymmetrical shape of the tracer profiles originated from the initial (Matano) planes. Indeed, the tracer profiles of $^{59}$Fe are significantly different on both sides of the Matano planes (set at $x = 0$ in Fig. 6). Such an asymmetry is seen especially strongly for the Fe/Fe-24Ga couple.

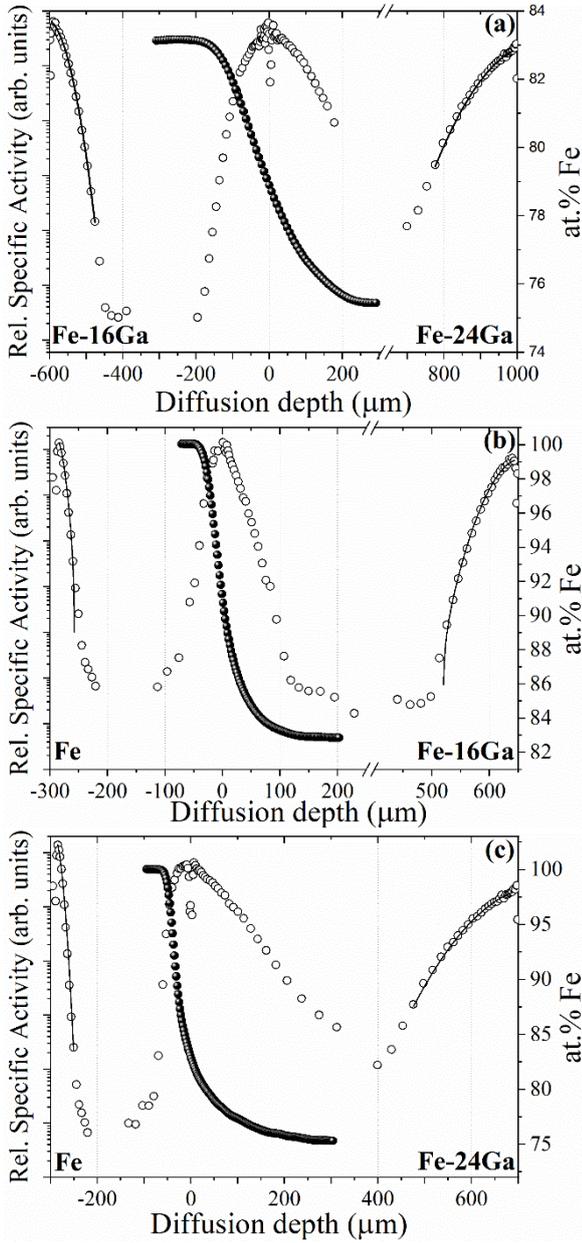

**Figure 6.** Penetration profiles of $^{59}$Fe tracer (circles, left and right ordinate) and the superimposed chemical diffusion profiles (spheres, right ordinate) in (a) Fe-16Ga/ Fe-24Ga, (b) Fe/Fe-16Ga, (c) Fe/Fe-24Ga diffusion couples. The chemical and tracer profiles are superimposed placing the origins of the respective abscise axes at the Matano plane determined by the chemical profile.

The tracer diffusion coefficients under chemical gradient are calculated using the theory proposed by Belova *et al.* [17–19],

$$D_{Fe}^{*} = -\left\{ \frac{V_m\left(\frac{\Delta x_{MT}}{V_m^0} + \int_{x_0}^{x}\frac{dx}{V_m}\right)}{2t} - \frac{1}{2Nt}\int_{N_0}^{N_{Fe}} x\, dN_{Fe} \right\} \bigg/ \left\{ \frac{d\ln C^*}{dx} - \frac{d\ln N_{Fe}}{dx} \right\}$$

(12)

Here $D_{Fe}^{*}$ is the composition-dependent tracer diffusion coefficient, $V_m^0$ is the molar volume at $x =$ 0, i.e., at the Matano plane. $C^*$ is the concentration of tracer that corresponds to the diffusion depth $x$ and $\Delta x_{MT}$ accounts for the difference in positions of the Matano planes for purely interdiffusion and combined tracer-interdiffusion couples.

From Eq. (12), it can be observed that the estimated tracer diffusion coefficients are functions of the diffusion depth. Especially, the relative positions of the Matano planes, $\Delta x_{MT}$, is crucial in determining the tracer diffusion coefficients of the respective diffusion couples. As was suggested [17], its values for different couples were determined by the condition that the tracer diffusion coefficients are continuous as a function of the depth.

The tracer diffusion coefficients obtained from the combined analysis of inter- and tracer diffusion data are presented in Fig. 7. The measured tracer diffusion coefficients are in perfect agreement with the independently measured tracer diffusion coefficients of the end-members and agree well with the literature data for Fe self-diffusion in α-Fe [40,41].

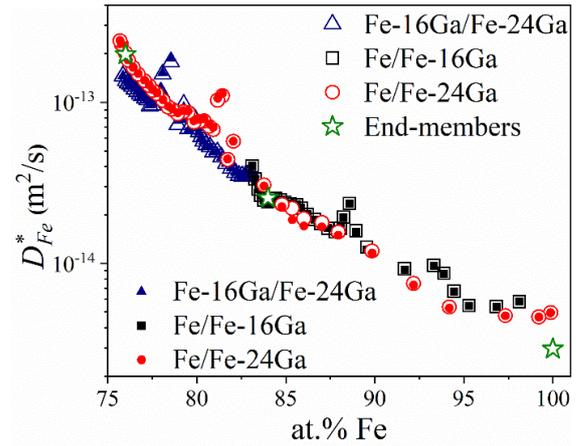

**Figure 7.** The Fe tracer diffusion coefficients as a function of the Fe concentration in Fe-Ga alloys after application of the combined analysis of chemical and tracer diffusion data. Plots show the comparison of the tracer diffusion coefficients with (open symbols) and without (closed symbols), taking into account the changes in molar volume. The tracer diffusion coefficients measured in independent experiments for the end-members are shown by stars.

The variation of the molar volume on the estimated tracer diffusion coefficients is found to be marginal, Fig. 7b. Again, we are attributing this fact to the almost linear dependence of the molar volume on the Fe concentration in the present alloys. Figure 7 shows the compiled tracer diffusion coefficients of all 3 couples. Though the three couples are measured independently, the results are in excellent agreement.

Figure 8 compares the Fe tracer diffusion coefficients, Eq. (12), and the interdiffusion coefficients, Eqs. (1) and (3). For subsequent estimates, the complete datasets for tracer and interdiffusion coefficients were fitted by second-order polynomial functions (in m²/s, for the composition interval of $1 \leq N_{Fe} \leq 0.7$ at $T = 1143$ K),

$$D^*_{Fe} = 3.6700 \times 10^{-15} \cdot \exp\{7.2999 \cdot (1 - N_{Fe}) + 35.4454 \cdot (1 - N_{Fe})^2\} \quad (13)$$

$$\widetilde{D} = 4.8352 \times 10^{-15} \cdot \exp\{17.4082 \cdot (1 - N_{Fe}) + 7.6917 \cdot (1 - N_{Fe})^2\} \quad (14)$$

and the results are given in Fig. 8b (solid and dashed lines, respectively).
Equations (8) could be written as:

$$\widetilde{D} = C_{Ga}V_{Ga}D_{Fe} + C_{Fe}V_{Fe}D_{Ga} = (N_{Ga}D^*_{Fe} + N_{Fe}D^*_{Ga})\Phi W_{AB} \quad (15)$$

Here the term related to the vacancy-wind effect for interdiffusion is expressed as $W_{AB} = 1 + \frac{2N_{Ga}N_{Fe}(D^*_{Ga}-D^*_{Fe})^2}{M_o(N_{Ga}D^*_{Fe}+N_{Fe}D^*_{Ga})(N_{Ga}D^*_{Ga}+N_{Fe}D^*_{Fe})}$ and $D^*_{Ga}$ is the tracer diffusion coefficient of Ga. The interdiffusion coefficient, $\widetilde{D}$, approaches the tracer diffusion coefficient of Ga in pure Fe, $D^*_{Ga}(N_{Ga} = 0)$, $N_{Ga} \rightarrow 0$, since simultaneously $\Phi \rightarrow 1$ and $W_{AB} \rightarrow 1$.

The thermodynamic factor $\Phi$ for the BCC phase in Fe-Ga system was also calculated using Thermo-Calc software [42] for the composition range 0-30 at.% Ga employing the Calphad description of the system provided by [43]. The calculated values of $\Phi$ at 1143 K can be approximated by the following polynomial function ($1 \leq N_{Fe} \leq 0.7$)

$$\Phi = 1.0 + 22.395 \cdot (1 - N_{Fe}) - 32.04 \cdot (1 - N_{Fe})^2 \quad (16)$$

The approximations (13), (14) and (16) allows an estimation of the Ga tracer diffusion coefficient using Eq. (15). The results are shown in Fig. 8b, dotted-dashed line.

On the other hand, the thermodynamic factors for the Fe-Ga system at the Kirkendall plane locations can be determined using Eq. (15) using the ratio of the intrinsic diffusion coefficients. Simultaneously, the Ga tracer diffusion coefficients can be determined, too, and the results are plotted in Fig. 8b, filled stars (the individual parameters are listed in Table 2). A good agreement of all results is seen. Nevertheless, the intrinsic diffusion data suggest that $D^*_{Ga}/D^*_{Fe} > 1$ at the concentrations corresponding to the positions of the Kirkendall planes, while inverse ratios are predicted using the available thermodynamic description. Note that the former were estimated without inclusion of any thermodynamic details. This fact substantiates requirements for further refinement of the thermodynamic description of the Fe-Ga system.

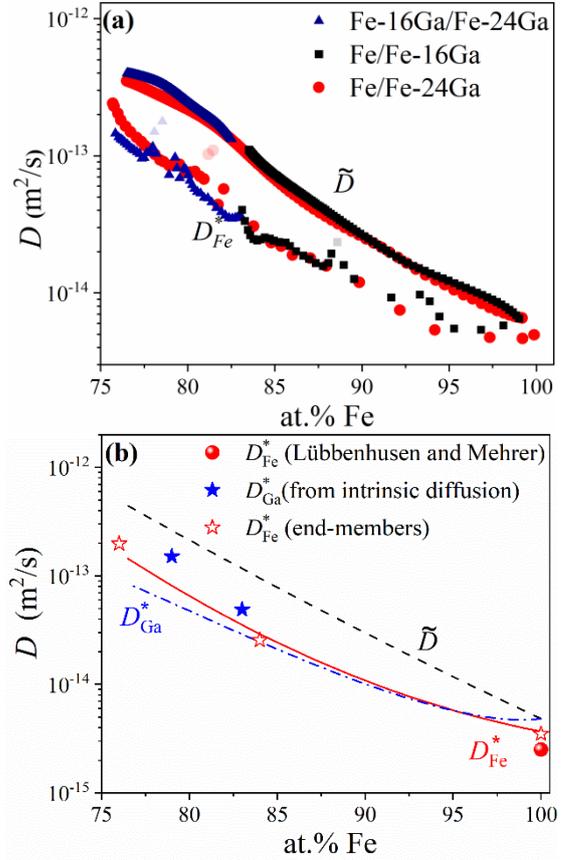

**Figure 8**. (a) Fe tracer and interdiffusion coefficients of the Fe-Ga couples. (b) Tracer diffusion coefficient of Ga, $D^*_{Ga}$ (dashed-dotted line), estimated using the Darken-Manning equation and the measured interdiffusion coefficient, $\widetilde{D}$ (dashed line), and the Fe tracer diffusivity, $D^*_{Fe}$ (solid line). The tracer diffusion coefficients of Fe measured for the end-members (open stars) are compared with the measurements of $D^*_{Fe}$ in pure α-Fe by Lübbenhusen and Mehrer [40] (sphere) and the estimated $D^*_{Ga}$ from the intrinsic diffusion coefficients (filled stars).

## 4. Conclusion

The combination of (radio) tracer source and interdiffusion couple experimental techniques is shown to allow a reliable determination of the composition dependence of tracer diffusion coefficients along the whole diffusion path. Diffusion in the (model) Fe-Ga system is investigated at 1143 K with the application of this novel technique using the ⁵⁹Fe radioisotope. The measurements are performed for the three couples Fe/Fe-16Ga, Fe/Fe-24Ga and Fe-16Ga/Fe-24Ga in overlapping concentration intervals, and excellent reproducibility of the results is found. A strong composition dependency of the interdiffusion coefficient and the tracer diffusion coefficient of Fe is established as well. The influence of the variable molar volume on the determined Fe tracer and interdiffusion coefficients is evaluated to be negligible for this system. The results confirm a high potential of the tracer-interdiffusion couple technique for producing highly accurate diffusion data. The estimation of the tracer diffusion coefficient of one component at the Kirkendall marker plane utilizing the tracer diffusion coefficients estimated directly following the radiotracer method is demonstrated, in which knowledge of the relevant thermodynamic details are not required. This has an immense benefit since reliable thermodynamic information are not available for various technologically relevant systems. As explained further, one can even estimate these parameters indirectly following this method.

This method is demonstrated in a binary system. If combined with the pseudo-binary method, one can determine the tracer diffusion coefficients of such elements as Al, Si, etc. (for which the tracer diffusion coefficients cannot be measured because of the absence of suitable isotopes) in the multicomponent systems even when the reliable thermodynamic functions are not available. Furthermore, the missing thermodynamic functions can be estimated by relating the intrinsic and tracer diffusion coefficients. This approach can be utilized to verify the reliability of thermodynamic descriptions established in various commercial thermodynamic databases by extending the details of binary and ternary system to the multicomponent system.

**Acknowledgments**. G.M.M. is grateful to the German academic exchange service (DAAD) for awarding a fellowship to conduct tracer diffusion experiments at the University of Münster, Germany. Financial support from the German Science Foundation (DFG), project DI 1419/11-1 is acknowledged. GEM and IVB acknowledge the support of the Australian Research Council, Discovery Project funding scheme (Project number: DP 170101812). ISG acknowledge the support of the Russian Scientific Foundation (Project number: 18-12-00283).